\documentclass[english,aps,twocolumn]{IEEEtran}
\usepackage[T1]{fontenc}
\usepackage[latin9]{inputenc}
\usepackage{graphicx}
\usepackage{cite}

\makeatletter

\makeatletter

\usepackage{esint}

\makeatletter

\DeclareFontEncoding{LGR}{}{}

\usepackage{soul}

\usepackage{xcolor}



\usepackage{babel}
\makeatother

\begin{document}

\title{Stability of $2\pi$ domain walls in ferromagnetic nanorings}

\author{\IEEEauthorblockN{
Gabriel D. Chaves-O'Flynn
\IEEEauthorrefmark{1}, 
Daniel Bedau
\IEEEauthorrefmark{1}, 
Eric Vanden-Eijnden
\IEEEauthorrefmark{2}, 
A. D. Kent
\IEEEauthorrefmark{1}, 
and D. L. Stein 
\IEEEauthorrefmark{1}\IEEEauthorrefmark{2}
}

\IEEEauthorblockA{\IEEEauthorrefmark{1}
Department of Physics, New York University, 4 Washington Place, New
York, New York 10003, USA}

\IEEEauthorblockA{\IEEEauthorrefmark{2}
Courant Institute of Mathematical Sciences, New York University, 251 Mercer Street, New
York, New York 10012, USA
}}

\maketitle
\begin{abstract}
The stability of $2\pi$ domain walls in ferromagnetic nanorings is
investigated via calculation of the minimum energy path that separates
a 2$\pi$ domain wall from the vortex state of a ferromagnetic
nanoring. Trapped domains are stable when they exist between certain
types of transverse domain walls, i.e., walls in which the edge defects
on the same side of the magnetic strip have equal sign and thus repel.
Here the energy barriers between these configurations and vortex magnetization states are obtained using the string method. Due to the
geometry of a ring, two types of $2\pi$ walls must be distinguished
that differ by their overall topological index and exchange energy.
The minimum energy path corresponds to the expulsion of a vortex.
The energy barrier for annihilation of a $2\pi$ wall is compared
to the activation energy for transitions between the two ring vortex
states. 
\end{abstract}

\section{Introduction}
Recent observations \cite{castao_metastable_2003,kunz_field_2009,klaui_head-to-head_2008}
in thin ferromagnetic stripes show magnetization configurations in
which the magnetization makes a full $2\pi$ turn in a localized region
of the stripe, while the rest of the stripe is magnetized parallel
to the edges of the stripe. These structures are stable against small
applied external magnetic fields which makes them potentially useful
for information storage devices\cite{muratov_bit_2009}. The same phenomenon 
has been observed in thin ferromagnetic annuli\cite{castao_metastable_2003,muratov_bit_2009}, which then results in 
the existence of a hierarchy of equally spaced metastable states\cite{chaves-oflynn_micromagnetic_2009}.

Reliable control of domain wall structures is crucial in the successful
design of magnetic nanodevices\cite{allwood_magnetic_2005,parkin_magnetic_2008}.
Here we explore the stability of $2\pi$ domain walls in ferromagnetic
nanorings using the string method \cite{e_string,e_energy_2003,e_simplified_2007}.
We find the energy barrier $\Delta E$ separating two metastable configurations. This barrier determines the escape rate from a metastable state through the Arrhenius law,
to leading order $e^{-\frac{\Delta E}{k_B T}}$\cite{hnggi_reaction-rate_1990}.
The string method also gives the minimum energy path and transition
state.

We distinguish two types of domain walls by their winding number in
the global (local) coordinate system $\omega$ ($\Omega$). We compare
the energy barriers that separates each type of wall from the ground
state to the activation energy that separates the two vortex configurations
\cite{martens_magnetic_2006,chaves-oflynn_micromagnetic_2009}.
A current flowing along the axis of the ring produces a circumferential
field. The application of this field has two effects.
First, the degeneracy of the two vortex states is lifted. Second,
the field determines the width of the domain wall; in its absence
the $2\pi$ walls could dissociate into two independent $\pi$ transverse
walls.

\section{Method}

Following previous work \cite{martens_magnetic_2006,chaves-oflynn_micromagnetic_2009},
we study a permalloy ring with the following dimensions and material properties: outer
radius $R_{2}=220$ nm; inner radius $R_{1}=180$ nm, thickness $t=2$
nm, magnetization saturation $M_{s}=8\times10^{5}A/m$, and exchange
length given by $l_{\mathrm{ex}}=\sqrt{\frac{2A}{\mu_{0}M_{s}^{2}}}=5.6$
nm. A current flowing along the axis of the ring produces a field $\mathbf{H}(\mathbf{r})=(hH_{c}(R_{1}+R_{2})/2r)\hat{\theta}$
A/m. Here $h=H/H_c$ and the characteristic
field strength at midradius is $\mu_{0}H_{c}=73.9$ mT (for $H>H_{c}$, the clockwise vortex
state is no longer stable). The calculations were performed at $h$=0.1.

Precessional effects do not modify the location of the critical points 
in the energy landscape: the exponential factor in the Arrhenius formula
 is unaltered if we ignore them. We consider the overdamped case so that 
 the escape trajectory follows
the negative gradient of the energy. This is done by integrating only
the damping term of the Landau-Lifshitz-Gilbert equation

\begin{equation}
\frac{d\mathbf{M}}{dt}=-\frac{|\gamma|\alpha}{M_{s}}\mathbf{M}\times(\mathbf{M}\times\mathbf{H_{\mathrm{eff}}}).\end{equation}
 Here $\alpha=1$ is the damping coefficient, $\gamma$ is the gyromagnetic
constant, and $\mathbf{H_{eff}}=-\nabla_{\mathbf{M}}E/\mu_{0}$ is
the effective magnetic field. The total micromagnetic energy $E$
is the sum of the exchange $E_{\mathrm{ex}}$, Zeeman $E_{\mathrm{Z}}$
and magnetostatic terms $E_{\mathrm{mag}}$.

The string method is necessary to calculate the minimum energy path between two
stable states ($\mathbf{M_{A}},\mathbf{M_{B}}$) when there is no a-priori knowledge of the transition state.
In practice, the path is discretized in N+1 images between $\mathbf{M_{A}}$ and
$\mathbf{M_{B}}$ denoted as $\mathbf{M}_{i}(t)\equiv\mathbf{M}_{i}(\mathbf{r},t)$ with
$i=0,...,N$. The images are updated using a two-step iteration procedure as follows: First, each image evolves using the publicly available
micromagnetic code OOMMF\cite{oommf} until the time reaches some
interval $\Delta t$ which we have selected to be 10 ps. This gives
a sequence of configurations:
\begin{equation}
\mathbf{M'}_{i}\equiv\mathbf{M'}_{i}(\mathbf{r})=\mathbf{M}_{i}(t)+\int_{t}^{t+\Delta t}\frac{d\mathbf{M}_{i}(t')}{dt}dt'\end{equation} Once all the $\mathbf{M'}_{i}(\mathbf{r})$ have been obtained, the
second step in the string method is a reparametrization step used to keep these images equidistant. First the complete
arc $s_{N}$ length of the trajectory is calculated by
\begin{equation}
s_{0}=0,s_{i}=s_{i-1}+|\mathbf{M'}_{i}-\mathbf{M'}_{i-1}|.\end{equation} The arc lengths are renormalized using $\alpha'_{i}=s_{i}/s_{N}$.
Finally we do a simple linear interpolation for all $i$ along the
trajectory so that \begin{equation}
\mathbf{M}_{i}(t+\Delta t)=\mathbf{M'}_{j(i)}+\frac{\mathbf{M'}_{j(i)+1}-\mathbf{M'}_{j(i)}}{\alpha'_{j(i)+1}-\alpha'_{j(i)}}(\frac{i}{N}-\alpha'_{j})\label{eq:interpolation}\end{equation} where $j(i)$ is the index of the string where $\alpha'_{j+1}\geq i/N\geq\alpha'_{j(i)}$.
During each step we observe the magnetic energy $E_{i}(t)$=$E_{\mathrm{ex}}(\mathbf{M}_{i}(\mathbf{r},t))+E_{\mathrm{Z}}(\mathbf{M}_{i}(\mathbf{r},t))+E_{\mathrm{mag}}(\mathbf{M}_{i}(\mathbf{r},t))$
as indicator of how far from convergence the current step is. The
iteration process is stopped when there is no visible change in the
function $E_{i}(t)$.

\section{Annihilation of $2\pi$ domain wall.}

\begin{figure}
\includegraphics[width=2.5in,angle=-90]{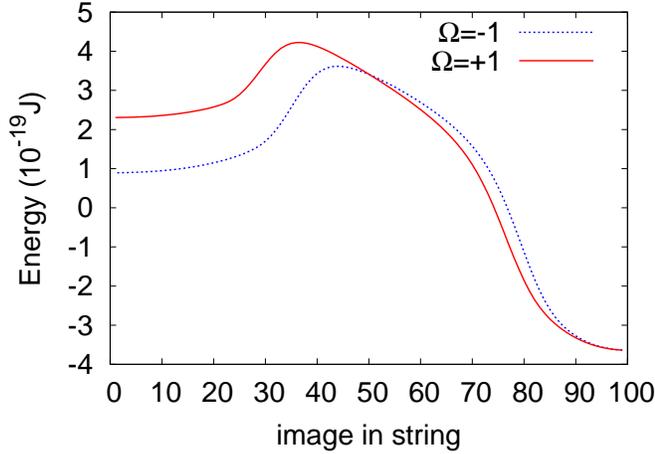}

\caption{\label{fig:annihilationof2piwall}Energy barrier for annihilation
of $2\pi$ domain wall under an external field $h=0.1$. Fig. \ref{fig:annihilationof2piwalls} shows the configurations for images i=0,50,100.}

\end{figure}

\begin{figure}
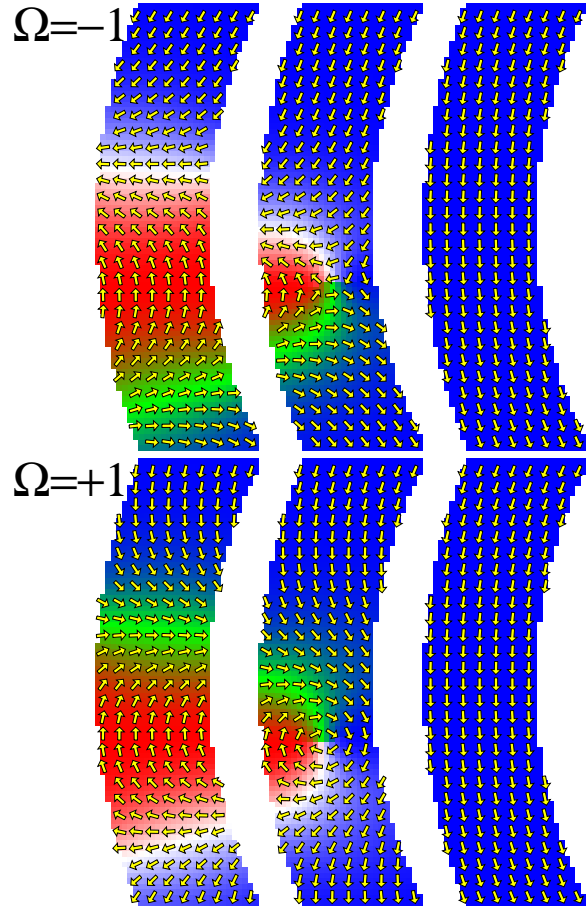

\includegraphics[width=3in]{FIG2}\smallskip{}
 \includegraphics[width=3in]{FIG3} 

\caption{\label{fig:annihilationof2piwalls}Segment of the ring encompassing each $2\pi$ domain wall. Minimum energy path for the annihilation
of $2\pi$ domain walls in consideration. (Above) $2\pi$ wall with
topological index $\Omega=-1$, (below) $2\pi$ wall with topological
index $\Omega=+1$. The configurations shown correspond to the following images in the string (left) i=0, (center) i=50, (right) i=100.}

\end{figure}

We now present the results of the string method to find the minimum
energy path for destruction of a $2\pi$ wall for the two types of
$2\pi$ domain wall, $\Omega=\pm1$. Fig. \ref{fig:annihilationof2piwall}
and Fig. \ref{fig:annihilationof2piwalls} present the
string energies and configurations after relaxation of the string.
The $\Omega=+1$, $2\pi$ wall decays into the counterclockwise configuration
by the expulsion of a vortex from the inner hole. On the other hand,
the $\Omega=-1$, $2\pi$ wall decays into the counterclockwise configuration
by the expulsion of an antivortex. This observation shows a
correspondence between a topological defect crossing the stripe and
the signature of the $2\pi$ walls being annihilated. %

For comparison purposes we provide magnitudes of the energy landscape
of ferromagnetic nanorings obtained with the string method. The lowest
energy barrier between the counterclockwise and the clockwise vortex
configurations passes through a configuration denoted as the instanton saddle \cite{martens_magnetic_2006}(with $\Omega=0$);
the activation energy of this event is $3.0\times10^{-19}$ J. This is
consistent with our previous work on nanorings \cite{chaves-oflynn_micromagnetic_2009}.
For a $2\pi$ wall with $\Omega=1$ the decay into the ground
state has an energy barrier equal to $1.9\times10^{-19}$ J. The annihilation of
a $2\pi$ wall with index $\Omega=-1$ has an energy barrier equal
to $2.8\times10^{-19}$ J. This shows the energy barrier to annihilate a $2\pi$
domain wall by the expulsion of a topological defect is comparable
to that of reversal between vortex states by a instanton fluctuation.

\section{Discussion}

For annuli with the dimensions being considered the magnetization
is constrained to lie in the plane of the ring; for the stable states
the magnetization can be considered to be independent of the radial
coordinate. The exchange energy $E_{\mathrm{ex}}$ of a stable state
is given by \cite{martens_magnetic_2006}:

\begin{equation}
E_{\mathrm{ex}}=\frac{\mu_{0}M_{s}^{2}tl_{\mathbf{\mathrm{ex}}}^{2}}{2}\ln\left(\frac{R_{2}}{R_{1}}\right)\left[2\pi(1+2\Omega)+\int_{0}^{2\pi}\left(\frac{\partial\phi}{\partial\theta}\right)^{2}d\theta\right]\label{eq:exchange}\end{equation}
 where $\phi(\theta)$ is the angle that the magnetization makes with
the tangent of the ring at a given angle $\theta$; $\Omega$ is the
``winding number'' of the magnetization with respect to the local
coordinate system.

The difference in winding numbers ($\Delta\Omega$) of the $2\pi$
walls considered results in an exchange energy difference $\Delta E_{\mathrm{\Omega}}$
between these two states. Using Eq. (\ref{eq:exchange}) the difference
can been shown to be approximately:

\begin{equation}
\Delta E_{\mathrm{\Omega}}[\mathbf{M}(\mathbf{r})]\approx2\pi\mu_{0}M_{s}^{2}tl_{\mathbf{\mathrm{ex}}}^{2}\ln\left(\frac{R_{2}}{R_{1}}\right)\Delta\Omega=1.298\times10^{-19}\ \mathrm{J}.\label{eq:exchangedif}\end{equation}
 The total exchange energy difference between these $2\pi$ walls
($\Omega=\pm1$) obtained from the micromagnetic simulation results
is $1.357\times10^{-19}\mathrm{J}$. Here we have ignored a contribution to this
difference of terms of the form $\int\left(\frac{\partial\phi}{\partial\theta}\right)^{2}d\theta$
since it is not a topological term. This shows a very good agreement
between an estimate obtained from the 1D model and the full numerical
simulation. The exchange energy  term is the biggest contributor to
the difference between the total energies of the two domain walls:
the numerical values from the demagnetization and Zeeman energy are
10 times smaller. The main point is that most of the energy difference
between these two types of $2\pi$ domain wall is the result of their
respective topological windings. It is worth noting that this is a curvature 
effect as can be seen from Eq. (\ref{eq:exchangedif}): the energy difference 
tends to zero in the limit when the radii approaches infinity --i.e. in the limit of a straight ferromagnetic strip.

We now consider the question of how to experimentally produce these two types of
$2\pi$ walls. We use the information contained in their global topological
number $\omega$ and compare it to other known states of nanorings.
In particular, the well-known "onion" state has $\omega=0$. Since
the onion is the remanent magnetization after saturation by an inplane
uniform field, one can produce the $\Omega=-1$ wall by applying a
strong field in-plane followed by a circumferential field. The two walls will approach and form
a $2\pi,\Omega=-1$ wall. Changing the direction of either the in-plane
field or the circumferential field will only change the final position
of the $2\pi$ wall, not its topological index.

The $2\pi,\omega=2$ domain wall cannot be produced using only uniform
and circumferential fields. However, we propose the following technique
to produce that configuration in nanorings: apply a strong dipolar
field ($\omega=2$) in the interior of the ring. This could be produced
by a small current loop with its axis coplanar to the structure, or
by bringing a magnetic tip close to the device. If this dipolar field
is strong enough, two transverse walls would be produced at opposite
sides of the ring; the magnetization vector at the centers of the
wall will point in the opposite direction of the overall magnetization
of the ring. Activating the circumferential field as the dipole strength
is decreased will result in the desired configuration.

One final question is whether resistance measurements can distinguish
the difference between the two types of $2\pi$ structures reported
in this work. For instance, one could attempt to use anisotropic magnetoresistance
effect to read the overall winding number of the configuration.
An estimate of this effect can be obtained by integrating $\mathbf{J}\cdot\mathbf{M}$
along a certain segment of the ring that spans the whole $2\pi$ domain
wall. Since $\mathbf{J}$ runs along $\hat{\theta}$ the AMR would
be proportional to $\langle M_{\theta}/M_{s}\rangle$ which can be
directly calculated from the $2\pi$ domain structure. For two electrodes
located at the top and bottom of the segment shown in Fig. \ref{fig:annihilationof2piwalls}
these values are 0.042 and 0.016 for the $\Omega=+1$ and $\Omega=-1$
walls respectively. It therefore should be possible to apply a current to probe the winding
number of the $2\pi$ domain.

\section{Conclusion}

We have presented results on the thermal annihilation of $2\pi$ domain
walls. We differentiated between two types of $2\pi$ walls through
their winding number in curved nanowires. We have observed a simple
arithmetical relation between the topological index of the different
configurations and the processes by which each structure decays into
the ground state. The fact that the energy difference between the
two states is dominated by the exchange energy allows to identify
the states through their winding number. The transition path requires
the motion of a singularity through the bulk: an antivortex destroys
$\Omega=-1$ walls, and a vortex annihilates $\Omega=+1$ walls. Similar
behavior is expected to work in linear stripes. Topological
defects are known to play a role in certain types of phase transitions;
here we have determined the mechanism by which they destroy $2\pi$
wall structures.

The two types of $2\pi$ domain wall correspond to distinct metastable
states: the greatest contribution to the energy difference comes from
the exchange energy difference. The energy can be directly associated
with the topological signature of the magnetization configuration.
The stability of these two states is comparable to the stability of
the clockwise configuration. We suggest to use AMR noise
measurements as a way to verify the presence of the two types of wall
in magnetic nanowires and study their stability. Thermal fluctuations should
generate both types of wall which could be recognized as two separate
values of the AMR.

Further micromagnetic exploration of the $2\pi$ wall annihilation
problem presented here can be done moving away from the overdamped
regime by using a nonzero precessional term. The results presented
here will be interesting to compare to this case. While we have presented results on a 2nm 
thick ring, we expect the observed transition states in thicker rings.
The key parameter is the ratio of the thickness to the mean radius 
which should be less than approximately 0.1 \cite{martens_magnetic_2006}. 
We plan to explore the energy barriers and transition states as a function of thickness and ring radii.

\section{acknowledgments}

This research was supported by NSF Grants No. NSF-DMR-0706522, NSF-PHY-0651077, NSF-DMS-0718172, 
NSF-DMS-0708140, ONR grant ONR-N00014-04-1-6046 and the NYU Dean's Dissertation Fellowship.

\bibliography{IEEEabrv,CF08}
\end{document}